# High Magnification Events by MOA in 2007

**Philip Yock**
*Department of Physics,*
*University of Auckland,*
*Auckland, New Zealand*
E-mail: `p.yock@auckland.ac.nz`

Gravitational microlensing events of high magnification provide exceptional sensitivity to the presence of low-mass planets orbiting the lens star, including planets with masses as low as that of Earth. The essential requirement for the detection of such planets in these events is that the FWHM of the light curve be monitored continuously, or as nearly continuously as possible.

The dependence of planet detectability on the magnification caused by microlensing, on the planet mass and planet location, and on the size of the source star, may be understood in terms of simple geometrical properties of microlensing that have been known since 1964. Planetary signals of low-mass planets are found to be approximately independent of the magnification caused by microlensing. This implies that planets can be detected in events over a wide range of magnifications, from moderately high values ~ 100 to very high values ~ 1000. The former values are likely to yield more clear-cut separations of the stellar and planetary features on the light curve, but they require larger telescopes to obtain precision photometry.

During 2007, twenty-four events with magnification exceeding 50 were detected by the MOA collaboration, of which about half were also detected by the OGLE collaboration. A quarter of the events received essentially continuous coverage of their FWHMs by follow-up collaborations, another quarter received partial coverage, and the remaining half received little coverage. Final analysis of these events is still underway, but casual inspection of the light curves reveals some possible planetary detections amongst them.

During 2008 it is hoped that fuller coverage of events of high magnification will be obtained with the addition of further telescopes to existing follow-up networks.

## 1. Introduction

Following observations and analyses that were made of the high-magnification gravitational microlensing event MACHO-98-BLG-35, the MOA collaboration concentrated its efforts on finding and analyzing further events of high magnification [1,2]. Although a clear planetary detection was not made in the above event, it served to alert the MOA group to the excellent sensitivity of the high magnification technique. In this report, the chief characteristics of 24 events with magnification greater than 50 that were found by the MOA group during 2007 are summarized. Analysis of these events is still underway. Very preliminary (eye-ball) results only are reported here. Some elementary theoretical considerations of the signals caused by low-mass planets in microlensing events of high magnification events are also given, and some observational plans for the future.

## 2. Perturbations by low-mass planets in high magnification events

Liebes discussed the general properties of high magnification microlensing events in 1964 [3]. He noted that, when the magnification is high, two arcs are formed, as depicted below.

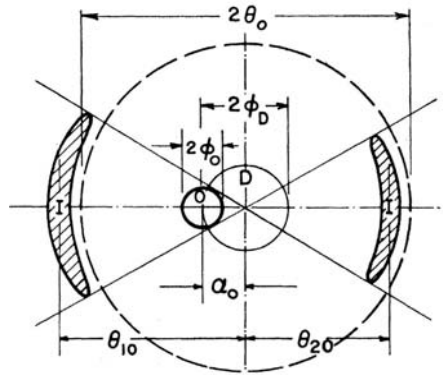

*Fig. 1. Einstein arcs formed at high manification, from [3].The width of the arcs is equal to the radius of the source star $\varphi_o$.*

Liebes showed that the widths and lengths of the arcs are simply determined by the size of the source star, and by the impact parameter of the event (denoted by $\alpha_o$ above) as depicted above. Remarkably, these simple relationships suffice to understand several of the main characteristics of planetary microlensing in these events as well.

During a microlensing event of high peak magnification, each arc slides around nearly half of the Einstein ring. If a planet is situated fairly close to the Einstein ring, it perturbs one or other of the arcs as it slides by, and thus betrays its presence on the light curve if the perturbation is sufficiently large.

Provided the planet is closer to the ring than the arc length, the duration of the planetary perturbation is given by the time required for the arc to slide past the planet. The width of the perturbation is therefore proportional to the radius of the source star. The height of the perturbation increases with the mass of the planet. In fact, it is proportional to mass. The fractional height of the perturbation is larger for smaller source stars, because these produce shorter arcs. In addition, the perturbation is larger for planets closer to the ring, provided they are not so close as to transit an arc. Finally, the fractional height of a planetary perturbation is approximately independent of the impact parameter and the peak magnification, because smaller impact parameters produce longer arcs that rotate faster.

The above expectations may be verified by simulations, using, for example, Dominik's code [4]. Consider first a "typical" event with $A_{max} = 100$, $\rho^* = 0.001$, $\theta = 60°$, $q = 10^{-4}$ and $b = 0.9$, where the symbols have their usual meanings ($\rho^*$ = angular radius of the source star relative to the angular radius of the Einstein ring, $\theta$ = angle between source star track and the lens axis, $q$ = planet:star mass ratio of the lens, and $b$ = projected planet:star separation relative to Einstein ring radius). The planetary perturbation for such an event is readily detectable, as shown below.

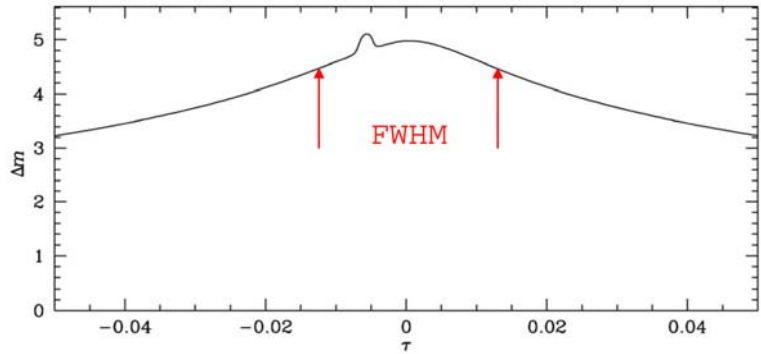

*Fig. 2. Readily detecatable planetary deviation for a low-mass planet with $q = 10^{-4}$, $A_{max} = 100$, $\rho^* = 0.001$, $\theta = 60°$ and $b = 0.9$. Most planetary perturbations occur within the FWHM of the light curve, thus necessitating dense coverage of this portion of the light curve only [5].*

Light curves with $q = 3 \times 10^{-4}$ and $3 \times 10^{-5}$ demonstrate that the height of the perturbation is proportional to planet mass q:-

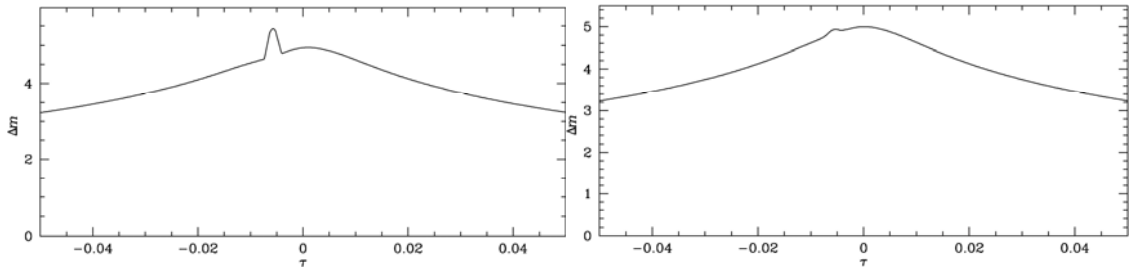

*Fig.3. Planetary perturbations for $q = 3 \times 10^{-4}$ and $3 \times 10^{-5}$ and other parameters as in Fig. 2.*

Light curves with b = 0.95 and 0.85 demonstrate that the perturbation decreases with the planetary distance from the Einstein ring:-

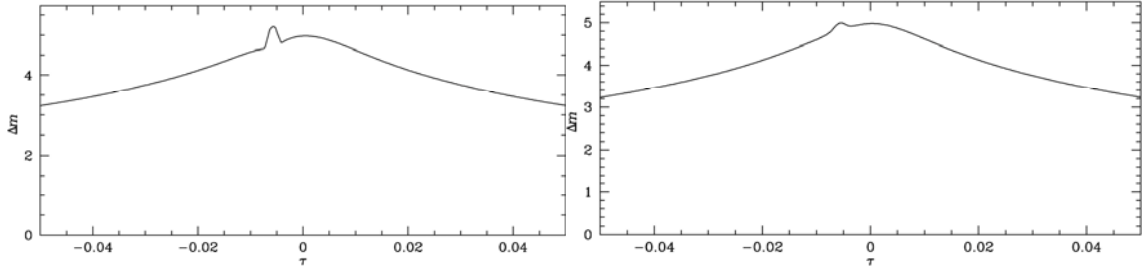

*Fig. 4 Planetary perturbations for b = 0.95 and 0.85 and other parameters as in Fig. 2.*

Light curves with $\rho^* = 0$ and 0.002 demonstrate that the height of the perturbation decreases as the size of the source star increases, but that the width of the perturbation increases:-

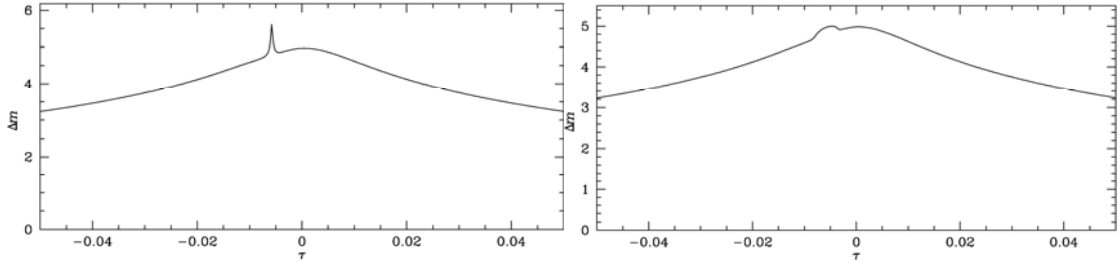

*Fig. 5. Planetary perturbations for $\rho^* = 0$ and 0.002 and other parameters as in Fig. 2.*

Finally, light curves with $A_{max}$ = 50 and 150 demonstrate that the perturbation is approximately independent of the magnification caused by microlensing:-

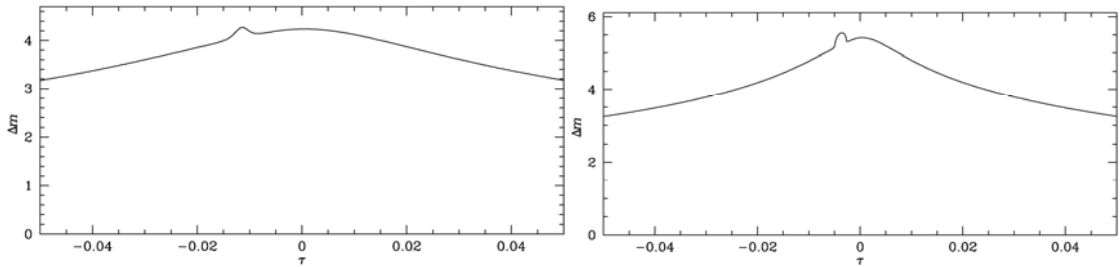

*Fig. 6. Planetary perturbations for $A_{max}$ = 50 and 150 and other parameters as in Fig. 2.*

The last result, namely that the planetary perturbation is (approximately) independent of magnification, is perhaps surprising. A qualitative understanding for this result was given above. However, as a further check on its validity, additional simulations were performed for $A_{max}$ = 100, 200, 500 and 2000 using different code [6]. This yielded similar results, as shown below in Fig. 7.

The impact parameters for the tests shown in Fig. 7 were 0.01, 0.005, 0.002 and 0, respectively. For the first three values, $A_{max}$ is given by its inverse, and it is seen that the planetary perturbation is (approximately) constant. For the last value, the finite size of the source star restricts $A_{max}$ to $2/\rho^* = 2000$ [3], and the planetary perturbation is reduced. In exceptional events where the impact parameter is smaller than the radius of the source star, the perturbative considerations given above do not apply.

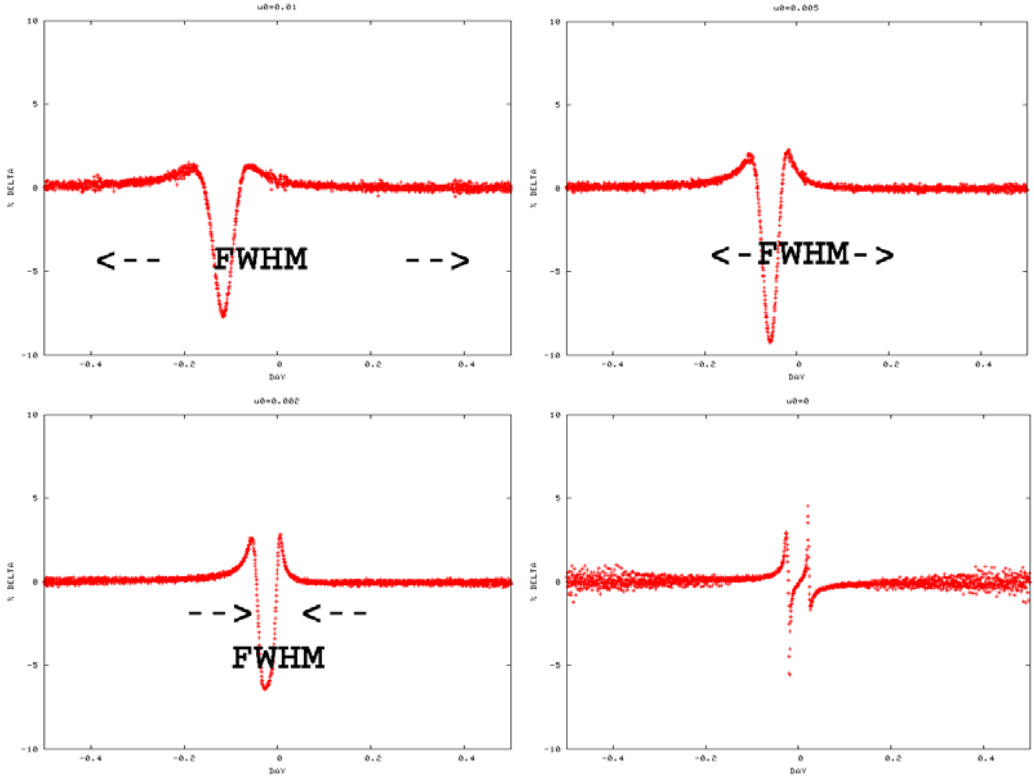

*Fig. 7. Planetary perturbations for $A_{max}$ = 100, 200, 500 and 2000, $q = 3 \times 10^{-5}$ and other parameters as in Fig. 2.*

The above figure demonstrates a further feature of planetary microlensing at very high magnification. The FWHM of the light curve gets 'squeezed' according to the relationship FWHM = $3.5t_E/A_{max}$ as the magnification increases [5]. The planetary deviation thus occupies a larger portion of the FWHM. For events with magnification of order a few hundred or less, it is likely that both the planetary and stellar portions of the light curve will be able to unambiguously recognized, and the stellar and planetary parameters of the event easily extracted. But for higher magnifications, it may be more difficult to disentangle the various features of the light curve.

The above light curves also demonstrate a further feature of planetary microlensing at high magnification, the presence of both positive and negative planetary perturbations. In Fig. 7 the lens axis relative to the source star track has been flipped through 180° relative to the previous examples. This reverses the perturbation. In general, one finds that the planetary perturbation is positive if the source star threads the lens components, and vice versa [5].

This completes the present discussion of planetary perturbations due to low mass planets in microlensing events of high magnification. Probably the main conclusions are that planetary perturbations depend sensitively on the size of the source star, so the latter needs to be known if precise results on planets are to be obtained, and also that planetary perturbations are only weakly dependent on the magnification produced by microlensing. It it thus not necessary to restrict searches to only those events with very high magnification. Indeed, events of moderately high magnification may be advantageous, by yielding clear-cut separations of the stellar and planetary features of the light curve. Any magnification greater than about 50 should be suitable, provided accurate photometry can be obtained over the FWHM.

Needless to say, giant planets may also be detected when the magnification caused by microlensing is high. However, many of the above considerations do not apply to giant planets. This is because the perturbations they produce are so large that they do not follow the simple perturbative arguments given above. In addition, giant planets may be detected at large distances from the Einstein ring. In such cases, the Einstein arcs will generally not be longer than the separation of the planet from the Einstein ring.

**3. High magnifcation events found by MOA in 2007**

The MOA telescope [7,8] operated successfully throughout 2007. Twenty-four events are listed at the MOA alert website with magnifications exceeding about 50 [9]. Approximately half of these events were found independently by the OGLE collaboration [10]. The events are listed in Table 1. In the following sections, brief discussion is given of a representative sample of them.

| MOA ID # | OGLE ID # | 50 < mag < 200 | 200 < mag | FWHM |
|---|---|---|---|---|
| 27 | 57 | * | | 0% |
| 33 | 114 | * | | 0% |
| 43 | - | | * | 20% |
| 51 | 13 | * | | 50% |
| 86 | 105 | * | | 0% |
| 103 | 050 | | * | 100% |
| 105 | 157 | * | | 0% |
| 110 | - | * | | 10% |
| 163 | 224 | | * | 100% |
| 170 | 220 | * | | 20% |
| 192 | - | | * | 10% |
| 193 | - | * | | 0% |
| 233 | 302 | * | | 60% |
| 281 | - | * | | 50% |
| 291 | - | * | | 0% |
| 312 | 388 | * | | 100% |
| 340 | 423 | * | | 100% |
| 378 | - | * | | 50% |
| 379 | 349 | | * | 100% |
| 397 | 538 | | * | 100% |
| 400 | - | | * | 100% |
| 403 | - | * | | 50% |
| 433 | - | | * | 0% |
| 464 | 514 | | * | 50% |

*Table 1. Chief characteristics of 24 high magnification events found by MOA in 2007. Identification nos. for those jointly found by OGLE are also listed. The final column denotes the approximate percentage of the FWHM that was densely photometered by the MOA, OGLE, MicroFUN [11] and PLANET/Robonet [12,13] groups combined.*

### 3.1 MOA-2007-BLG-027/OGLE-2007-BLG-057

This was the first event of high magnification to be found in 2007. It peaked in mid March at which time it is difficult to obtain continuous coverage of events that occur in the galactic bulge. The peak magnification was approximately 60, but no data were obtained by MOA or any other collaboration during the FWHM. The light curve for the event as recorded by the MOA telescope is shown in Fig. 8. The data were obtained by the "difference imaging" technique [14]. This yields photometry with accuracy approaching the statistical limit in crowded stellar fields. The data shown for this and subsequent events were posted at the MOA alert website in real time to enable follow-up observatories to be carried out around the globe.

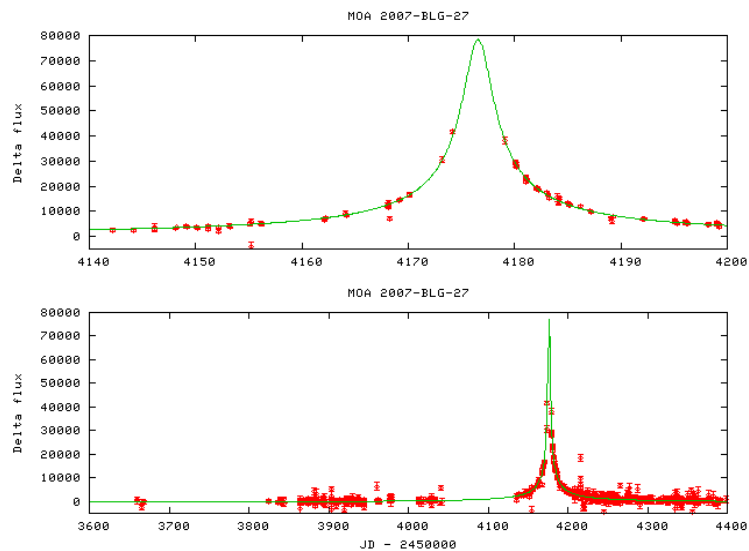

*Fig. 8. Photometry of MOA-2007-BLG-27 from the MOA alert website [9].*

### 3.2 MOA-2007-BLG-105/OGLE-2007-BLG-157

The light curve for this event for which $A_{max}$ was ≈ 60 appears below. This event displays features that are not uncommon. The absence of data on the night before the peak (JD2454205) precluded an alert of impending high magnification being issued. Despite the absence of an alert, the FWHM of the event was quite well covered by the MOA telescope. Nine measurements are shown above on the night of the peak. These were made as part of the normal MOA observational strategy, which includes very dense sampling of a few selected fields quite near the Galactic centre, and less dense sampling of more distant fields. The above event occurred in one of the latter fields. Casual inspection of the data would appear to argue against the presence of planets with $q \sim 10^{-4}$ at projected separations $0.9 < b < 1.1$, as perturbations comparable to that shown in Fig. 2 appear to be absent. However, the paucity of the data on the night of the peak precludes a firm conclusion being drawn without a full analysis of the data being undertaken.

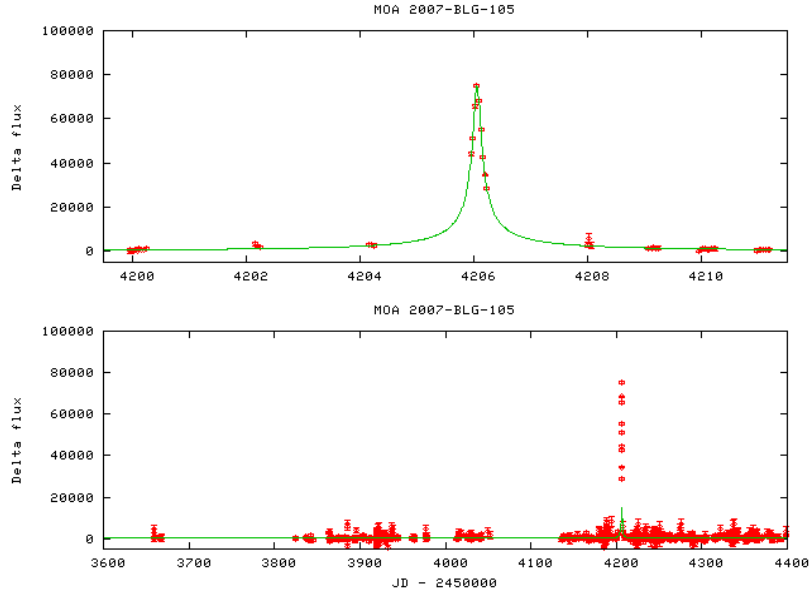

*Fig. 10. Light curve for MOA-2007-BLG-105 from the MOA alert website.*

### 3.3 OGLE-2007-BLG-224/MOA-2007-BLG-163

The light curve for this event exhibits the way that co-operation between survey and follow-up groups is hoped to work. The peak of the event was alerted ahead of time, and the FWHM was successfully photometered by the MicroFUN and OGLE collaborations. The light curve recorded by MOA is shown below.

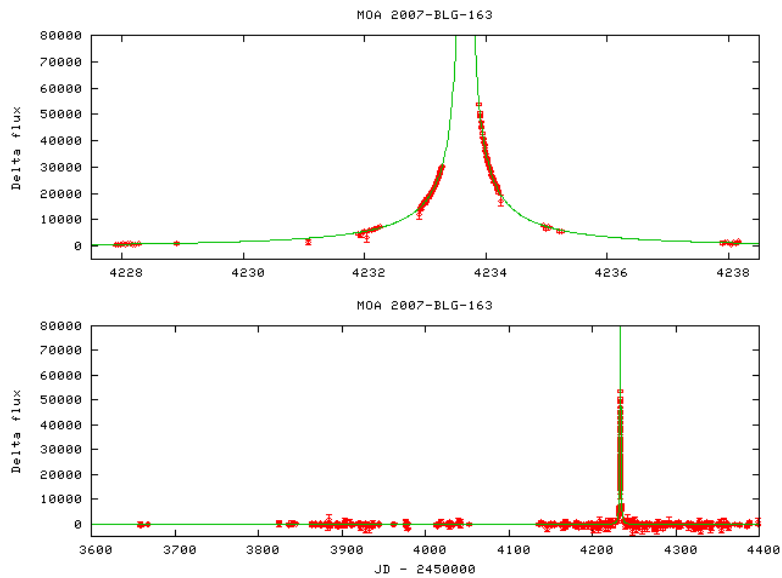

*Fig. 11. Light curve of MOA-2007-BLG-163 recorded by MOA.*

### 3.4 MOA-2007-BLG-176

Fairly good coverage of the peak of this event was obtained by the MOA collaboration, and further coverage was obtained by MicroFUN. The MOA data clearly show the rounded peak caused by the finite size of the source star. Casual inspection of the light curve suggests the absence of planets with $q \sim 10^{-4}$, $0.85 < b < 1.15$ and $70° < \theta < 110°$ or $160° < \theta < 200°$.

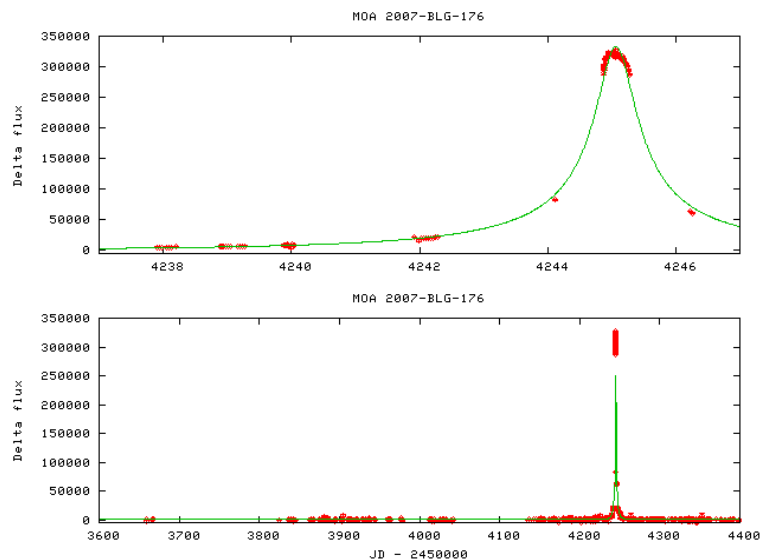

*Fig. 12. Peak light curve of MOA-2007-BLG-176 [13].*

### 3.6 MOA-2007-BLG-192

This light curve of the peak of this event is shown below. The event is qualitatively similar to MOA-2007-BLG-105 shown above, in that the coverage during the FWHM is not dense. Moreover, it is also not complete. Despite these shortcomings, there appears to be a clear discrepancy form the light curve for a single lens. A possible downward excursion is seen at JD ≈ 2454245.3 that could be fit with a planet with q ~ $10^{-4}$, b ~ 0.9 and θ ~ 60°. Further analysis of these data would be required to draw a firm conclusion.

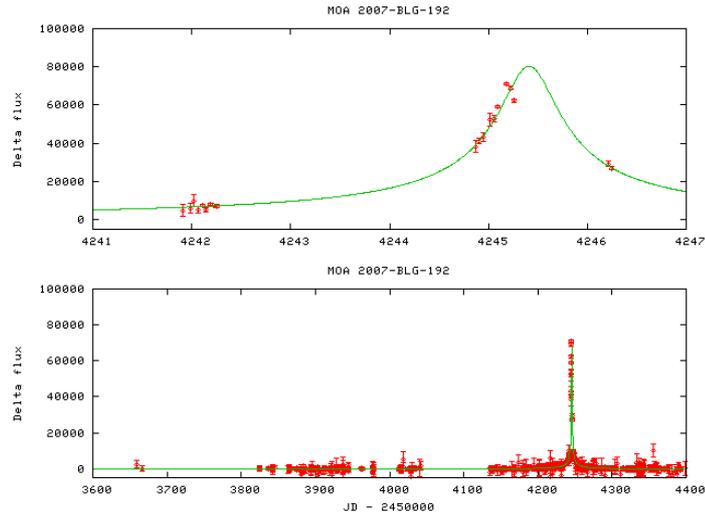

*Fig. 13. The light curve of MOA-2007-BLG-192 by the MOA group. .*

### 3.7 MOA-2007-BLG-312/OGLE-2007-BLG-388

The peak of the light curve, which peaked over New Zealand, is shown below.

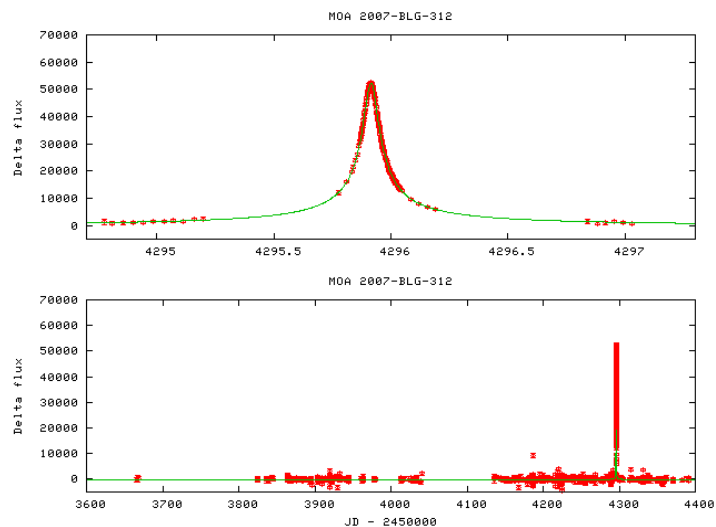

*Fig. 14. Light curve of MOA-2007-BLG-312 recorded by the MOA group.*

The apparent absence of planetary excursions during the FWHM suggests the absence of planets with q ≈ $10^{-4}$ and 0.85 < b < 1.15, but, as in all cases, a full analysis would be required to draw a firm conclusion.

### 3.8   OGLE-2007-BLG-349/MOA-2007-BLG-379

The light curve for this event as shown below appears to exhibit a large deviation from the single-lens light curve. However, the incompleteness of the light curve obtained by the MOA group would preclude an unambiguous interpretation being drawn. Fortunately, the FWHM of this event was fully monitored through the combined efforts of the world-wide microlensing community. This should yield a unique interpretation of the data in the future.

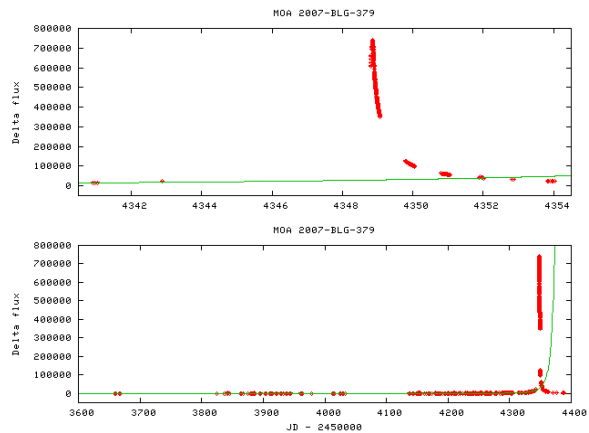

*Fig. 15. Light curve by MOA of MOA-2007-BLG-379*

### 3.9   MOA-2007-BLG-397/OGLE-2007-BLG-538

The light curve for this event is comparable to that of MOA-2007-BLG-312 shown

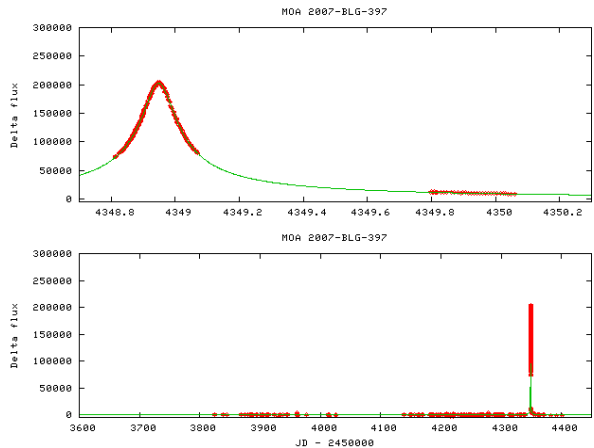

*Fig. 16. Light curve of MOA-2007-BLG-312*

above. A similar preliminary conclusion on the absence of planets could be drawn.

### 3.10   MOA-2007-BLG-400

The light curve for this event is shown below. The peak of this event was fully monitored by the MicroFUN and OGLE groups from Chile.

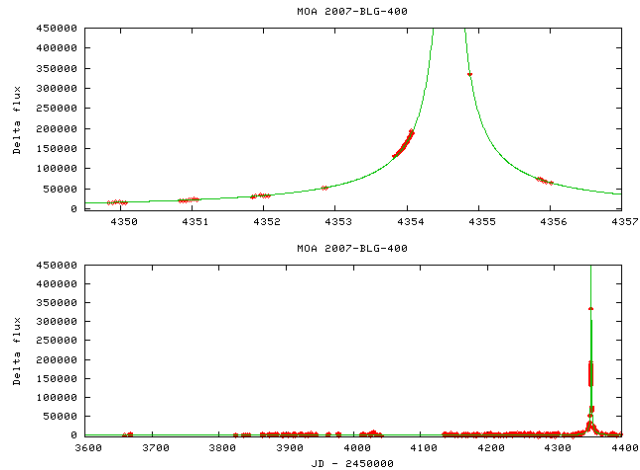

*Fig. 17. Light curve of MOA-2007-BLG-400*

### 4.   New telescopes

Despite the heroic efforts of follow-up groups to obtain complete coverage of the FWHM of every event with high magnification that was detected in 2007, the vagaries of the weather rendered the task impossible with the telescopes that were available. To this end, efforts were made during the year to attract the interest of further observatories in new locations, with some success. During 2008, it is hoped that the armada of telescopes operated by the MicroFUN, PLANET and Robonet collaborations will be augmented with further strategic telescopes located in Taiwan (Lulin Observatory), Hawaii (Mauna Kea), Namibia (ATOM telescope), Texas and South Africa (MONET network) and Blenheim (New Zealand) [15-19]. In addition, the results of test observations made from Dome A in Antarctica will be eagerly awaited [20].

### 5.   Conclusions

The hunt for planets by gravitational microlensing has been in a state of sustained growth during the last few years due to the dedicated and inspired efforts of members of all the participating collaborations. Exciting results have been obtained in this period, and it is guaranteed that further exciting results will unfold in the near future on low-mass planets orbiting stars in our galaxy. Analysis of the high-magnification events described above, and other events not described here, is certain to yield useful information on planetary statistics and on planetary formation processes.

**Acknowledgments**

The author is deeply indebted to many friends in several collaborations who made the above described work both possible and a pleasure. The light curves shown in this paper are testament to the skills and efforts of all members of the MOA collaboration. Financial support by the Angles Astrophysics Network for Galaxy Lensing Studies, the Marsden Fund of New Zealand and the University of Auckland is gratefully acknowledged.